# MICROWAVE FLASH SINTERING OF METAL POWDERS: FROM EXPERIMENTAL EVIDENCE TO MULTIPHYSICS SIMULATION

Charles Manière[a], Geuntak Lee[a,b], Tony Zahrah[c], Eugene A. Olevsky[a,d]*

a) Powder Technology Laboratory, San Diego State University, San Diego, USA
b) Mechanical and Aerospace Engineering, University of California, San Diego, La Jolla, USA
c) MATSYS Inc., Sterling, USA
d) NanoEngineering, University of California, San Diego, La Jolla, USA

**Abstract:** Flash sintering phenomena are predominantly associated with ceramics due to thermal runaway of their electric conductivity noticeably represented in materials such as zirconia or silicon carbide. Because of their high electric conductivity, flash sintering of metals is nearly inexistent. In this work, an original metal powder flash sintering method based on a microwave approach is presented. Within the developed approach, an unusually fast (60 s) thermal and sintering runaway of Ti-6Al-4V powder is experimentally revealed under microwave illumination. This phenomenon is simulated based on an electromagnetic-thermal-mechanical (EMTM) model. The developed multiphysics model reveals that the metal powder specimen's runaway does not result from its intrinsic material properties, but results from the resonance phenomenon thermally activated by the surrounding tooling material. The EMTM simulation predicts with a very good accuracy the microwave repartition and the resulting densification and powder specimen's shape distortions observed experimentally. The comparison of the microwave and conventional sintering kinetics indicates an important acceleration of the sintering behavior under microwave heating. The developed sintering approach has a potential of the implementation for time-effective mass production of small metal parts.

---

* Corresponding author: EO: Powder Technology Laboratory, San Diego State University, 5500 Campanile Drive, San Diego, CA 92182-1323, Ph.: (619)-594-6329; eolevsky@mail.sdsu.edu

## 1. Introduction

Since the publication of Cologna *et al*. [1] in 2010, the flash sintering phenomena encounter a great interest in the field of sintering studies. Flash sintering is considered to be one of the most efficient sintering approaches in terms of energy consumption, operating time and productivity for acceptable microstructures. The traditional flash sintering approach [1–4] involves resistive Joule heating of a green sample between two electrodes. In this process, the first incubation stage is observed where the temperature is slowly increased. In the second stage, a thermal runaway happens where the temperatures are quickly increased and accompanied by the fast sintering of the specimen in less than a minute. The thermal runaway profile of this process can be attributed to the negative temperature coefficient (NTC) behavior of the material [5–7]. The materials classically employed for flash sintering, such as zirconia or silicon carbide, require a preheating in a furnace to activate the electric conductivity of the green specimen, which is too low at room temperature [2].

Different alternative approaches have been developed based on the spark plasma sintering (SPS) [8–10], plasma electrode sintering [11], pressure-assisted flash sintering [12], and microwave sintering [13,14]. The flash spark plasma sintering (FSPS) approach using graphite felt as a preheating element has been employed to fully consolidate $ZrB_2$ [9] and SiC [10] during 17-40 seconds. Another flash spark plasma sintering (ultra-rapid hot pressing) approach using a copper sacrificial tooling element has been employed and allowed the full densification of SiC in 1 second [8]. A contactless flash sintering based on arc plasma electrode has been also developed enabling the full densification of carbide materials (pure $B_4C$ and $SiC:B_4C$ 50wt%) in about a few seconds of the discharge time [11]. Under microwave illumination, NTC ceramics heating demonstrates hot spot instabilities and exhibits a similar thermal runaway [15,16]. The microwave power dissipation of dielectric ceramics depends on the permittivity imaginary part that can be expressed by the material electric conductivity [17,18]. Therefore, for similar reasons, the NTC behavior of some ceramic materials responsible for thermal runaway in traditional resistive flash sintering approaches can also generate a thermal runaway under microwave illumination [13,15]. The microwave flash sintering of ceramics has been demonstrated for $Al_2O_3$, $Y_2O_3$, $MgAl_2O_4$, and $Yb:(LaY)_2O_3$ with dense (98-99%) microstructures [13,19,20]. These studies revealed by SEM and AFN characterization the presence of a particle surface softening/melting mechanism that significantly accelerated the sintering for $Yb:(LaY)_2O_3$ [13]. For the same material, transparent specimens have been obtained by the microwave flash method [14]. One of the microwave flash sintering advantages is the volumic and contactless heating of the sample; the flash process can then be entirely controlled by the microwave power.

Most of the materials employed for flash sintering described in the literature are oxides or carbides. A recent flash sintering review [21] estimates the proportion of different materials used for flash sintering as 96% for oxide, carbide and only 3% for conductive high temperature ceramics like $MoSi_2$, $ZrB_2$. The presence of common metals and alloys among the materials subjectable to flash sintering is nearly inexistent, because their resistivity properties do not manifest the NTC behavior and, in turn, no natural thermal runaway under constant applied voltage. The exception is the flash microwave inkjet techniques of polymer/silver mixtures under antenna microwave illumination that allow fast agglomeration and consolidation of silver particles [22,23].

In this paper, we show that a thermal runaway accompanied by a fast densification similar to the traditional flash sintering method can occur also in metal powders when using a microwave heating approach. It is shown experimentally and demonstrated theoretically how an unusual metal thermal runaway can be obtained based on a resonance phenomenon thermally activated by the surrounding tooling materials. In the present study, the experimental results have been compared to the outcomes of the electromagnetic-thermo-mechanical (EMTM) simulation able to predict the cavity microwave distribution, heating, sample densification and shape distortions [24]. In addition, the conventional sintering densification behavior has been compared to the microwave sintering experiments and showed a considerable densification kinetic acceleration under the metal powder flash microwave sintering mode.

2. Experimental procedures

The microwave sintering experiments have been carried out using the microwave furnace Bloomden PMF15B. The microwave cavity is made of a rectangular waveguide where a magnetron (WITOL 2M343K E625) is connected and a 135 mm diameter cylindrical heating area is located (see cavity fig.1). Two SiC susceptors and zirconia balls bed are employed in the heating area. The effective heating area dimensions are 50 mm diameter and 30 mm height, the other areas of the 135 mm cylindrical cavity are filled with fibrous alumina-silica insulation (80%$Al_2O_3$-20%$SiO_2$). The samples are pre-consolidated by spark plasma sintering (SPSS DR. SINTER Fuji Electronics model 5015) to a relative density of 40% in order to avoid using a polymeric binder. The sample dimensions after pre-consolidation are 10 mm of diameter and 7 mm height. Ti-6Al-4V powder including 50 μm agglomerates has been employed (see in fig.2a the SEM image of the powder). The energy dispersive x-ray spectroscopy (EDS) analysis reveals the presence of a certain level of powder particle surface oxidation (fig.2b). The microwave sintering has been conducted utilizing two heating modes. On the one hand, the mode when the Ti-6Al-4V sample is located on the zirconia bed in the center and at the edge of the heating area (called "free surface" FS mode) has been used. On the other hand, the mode when the sample is located in an alumina container and surrounded by a 45-65 nm SiC nano-size powder (called "powder surrounded" PS mode) has been employed. The microwave sintering tests were performed using the constant applied power of 1300 W.

The microwave sintering outcomes have been compared to the results of the conventional sintering using a dilatometer (Unitherm model 1161 dilatometer system) with a cycle of 10 K/min heating rate up to 1480°C. Both microwave and dilatometer test were conducted under argon atmosphere in order to prevent oxidation of the samples. The sample temperature has been estimated by an S type thermocouple for the dilatometer and by a pyrometer (Mikron infrared MI-P140, temperature range 200-1300°C) for the microwave sintering. In order to reveal the Ti-6Al-4V sample microstructures, the specimens where cut in half, mirror polished and etched using the Shelton mixture [25] (HF 10ml, $HNO_3$ 5 ml, $H_2O_2$ 30 ml, $H_2O$ 10ml).

**3. Theory and calculations**

*3.1 Electromagnetic-thermal-mechanical simulation*

The analysis of the microwave heating and sintering behavior requires a comprehensive multiphysics simulation including an electromagnetic part to describe the wave distribution in the cavity, a thermal part for the temperature field repartition, and a mechanical part for the resulting sintering of the sample. Based on our previous work [24], a fully coupled electromagnetic-thermal-mechanical (EMTM) simulation has been employed. This model also includes advanced simulation tools to calculate the surface-to-surface thermal radiation in the internal hot areas. The governing equations of the EMTM model are described in Refs. [15,24]. The external metallic surfaces are subjected to a surface-to-ambient thermal radiation and the convective heat loss is described by an emissivity of 0.79 and the convection heat flux of 5 $W.m^{-2}.K^{-1}$ [26]. The internal boundary emissivity and the electromagnetic-thermal-mechanical material properties are reported in Table A [15,24,27–32].

*3.2 Material parameters determination*

Some material properties depend on porosity and temperature and require a theoretical approach to be estimated. The dielectric properties of the porous tools like the zirconia balls bed and the nano-SiC powder have been estimated by the effective medium approximation equation (1) [15,18].

$$C_S\left(\frac{\varepsilon_{solid}-\varepsilon_{eff}}{\varepsilon_{solid}+2\,\varepsilon_{eff}}\right) + (1-C_S)\left(\frac{\varepsilon_{gas}-\varepsilon_{eff}}{\varepsilon_{gas}+2\,\varepsilon_{eff}}\right) = 0 \qquad (1)$$

with, $\varepsilon_{solid}$ the solid phase complex permittivity (zirconia or silicon carbide), $\varepsilon_{gas}$ the gas permittivity equal 1, $\varepsilon_{eff}$ the effective complex permittivity of the whole porous material to be determined, and $C_s$ the solid concentration that is associated with the relative density.

The thermal conductivity of the zirconia balls bed and nano-SiC powder can be determined by another equation (2) that also originates from the effective medium approximation and includes the thermal conductivity of the material and argon gas [18].

$$2\kappa_{eff}^2 + [\kappa_{solid}(3\theta-2) + \kappa_{gas}(1-3\theta)]\kappa_{eff} - \kappa_{solid}\kappa_{gas} = 0 \qquad (2)$$

where $\kappa_{solid}$ is the solid phase thermal conductivity (zirconia or silicon carbide), $\kappa_{gas}$ the gas thermal conductivity (argon here), $\kappa_{eff}$ the effective thermal conductivity of the whole porous material to be determined, and $\theta$ is the porosity.

The non-compacted nano-SiC powder has a very high porosity level (95.2%), and the consequence of this is a very low dielectric dissipation represented by the permittivity imaginary part (0.004 at 300 K and 0.009 at 1700 K) that does not impact the heating of the sample even despite silicon carbide is a very good susceptor at full density (0.3 at 300 K and 2 at 1700 K). The other consequence is an extremely low thermal conductivity (about 0.05 W.m$^{-1}$.K$^{-1}$). If we consider a sample surrounded by the nano-SiC powder in PS mode, the heat generated in the sample will be easily confined in it.

The determination of the Ti-6Al-4V sample electromagnetic properties is a difficult task because metals can reflect microwaves and, in certain circumstances, generate sparks under microwave illumination. In a powder state, metals can be heated under microwave radiation and also sintered to full density [33]. The explanation of this phenomenon is still debatable but it is broadly accepted that the microwave penetration is higher in metal powders compared to fully dense metals [17,18,33]. A significant microwave dissipation is also attributed to the formation of Eddy currents generated because of the microwave alternative magnetic field applied to the metal particles [18,34]. Based on these two widely recognized phenomena, the microwave complex permittivity and permeability have been estimated. As suggested by Rybakov *et al.* [18] the permittivity of the metal powder can be determined by the effective medium approximation that considers the porosity of the metal particles and the surrounding oxide shell (see eq (3)). The permeability $\mu$ can be also estimated theoretically based on the Eddy current contribution [34] (see eq (4)).

$$2\beta\varepsilon_{eff}^2 + [\beta(3C_s - 2) + \alpha\varepsilon_{TiO_2}(1 - 3C_s)]\varepsilon_{eff} - \alpha\varepsilon_{TiO_2} = 0 \quad (3)$$

Where $\varepsilon_{TiO_2}$ is the TiO$_2$ oxide shell permittivity [35], $\alpha = 2\varepsilon_{TiO_2} + \varepsilon_{Ti-6Al-4V} + 2(\varepsilon_{Ti-6Al-4V} - \varepsilon_{TiO2}/\zeta 3$, $\beta=2\varepsilon TiO2+\varepsilon Ti-6Al-4V+\varepsilon TiO2-\varepsilon Ti-6Al-4V/\zeta 3$, $\zeta=1+t/r$, r is the alloy particle radius, $t$ the oxide shell thickness, and $\varepsilon_{Ti-6Al-4V}$ the alloy permittivity that can be related to the conductivity [17].

$$\mu = 1 - \frac{3}{2}\left(1 - \frac{3}{k^2r^2} + \frac{3}{kr}cotkr\right)C_s \quad (4)$$

with $k = (1 + i)/\delta$, $\delta = c\sqrt{2\varepsilon_0/(\sigma\omega)}$ the skin depth, $c$ the light speed, $\sigma$ the alloy electrical conductivity, the angular frequency, $\varepsilon_0$ the vacuum permittivity. All the calculations have been conducted assuming 2.45GHz microwave frequency and with similar conditions to those used by Rybakov *et al.* [17]. The temperature and relative density dependence of the Ti-6Al-4V powder are reported in fig.3.

## 4. Results and discussion

*4.1 Preliminary microwave heating tests*

Ludo *et al*. [36] showed that Ti-6Al-4V powdered samples can be direct microwave heated but the heating is unstable and not reproducible. The authors showed that a stable heating with heating rate of about 30 K/min up to temperatures allowing the full sample sintering was possible using a SiC susceptor ring. Based on this, we first tried the microwave metal powder heating using the hybrid FS mode. The results reported in fig.4a show that both in the center and edge locations the sample heating is initially very fast and about 80 K/min, but spark phenomena happen at about 600-700°C. When the sparks appear, the microwave radiation is turned off which prevents the heating to sintering temperatures. Contrary to Luo *et al.* [36], the used in the present study cavity and the microwave furnace configuration do not allow the sintering of the sample by a simple hybrid heating configuration. To prevent the formation of sparks, another approach has been chosen where the sample is surrounded by a nano-SiC powder, that is - the above-mentioned PS mode.

*4.2 Microwave metal flash heating evidence*

The temperature profiles for all the PS mode tests are reported in fig.4b. The first attempt (red color curve in fig.4b) reveals an unexpected ultra-fast heating. In about 5 min of applied microwave power, the alumina container was fractured by thermal stresses, and the Ti-6Al-4V sample melted. It is to be noted that in the PS mode, the pyrometer does not indicate the sample temperature but, instead, the alumina container temperature. The temperature gap indicated by the red color curves in fig.4b originates from the transition of the container temperature to the sample temperature when the alumina container was broken. This first attempt suggests that an intermediate stage, when the sample is consolidated but not melted, is possible. Therefore the initial sintering strategy was modified in favor of controlling the flash heating. In order to obtain a sample consolidated but not melted, the microwave furnace was turned off at earlier heating times. First, the microwave furnace was turned off at 300°C (container temperature) and at that time no change in the sample dimensions has been observed, meaning that no sample densification occurred. When the microwave heating was shut down at 450°C, a significant reduction of the sample dimensions without apparent melting or container breakage has been observed. These tests show that the sample heating resulting in sintering is possible and occurs very briefly. In less than 2 minutes, the sample is heated to temperatures allowing sintering, and in one additional minute the sample is melted. The microstructure and the EMTM simulation of the sample densification are analyzed in the following sections.

*4.3 Flash microwave sintered microstructures obtained*

The flash sintered sample exposes a truncated cone shape distortion. The optical microscopy revealed 80% densified microstructure (fig.5) at the top, and a transition in the center to the 50% dense microstructure at the bottom part of the sample. The transition in the center is about 20 µm large and looks like a densification front starting from the top and propagating to the bottom of the sample. These facts suggest a higher temperature at the top area of the sample. The desirable α+β lamellar microstructure is observed at the sample's top with α lamella thickness between 1-5 µm. The ultra-fast heating apparently reaches far higher temperatures than the one corresponding to the β phase transition (980°C), and the low thermal conductivity of the surrounding nano-SiC powder prevents a fast sample cooling allowing the α phase lamella to sufficiently grow into the β grains avoiding the martensitic microstructure [37]. As a result, the flash sintered sample microstructure is porous but possesses an interesting lamellar structure. Further improvement of the densification of the sample requires a better control of the sample temperatures. In order to explain the ultra-fast heating of the sample as well as the observed microstructures, the EMTM simulation of the process is described in the next section.

*4.4 Metal powder microwave thermal runaway simulation*

To understand the heating characteristics of the Ti-6Al-4V powder, the electromagnetic properties of the alloy are fist simulated in a simple waveguide configuration and compared to the typical behavior of 3Y-ZrO$_2$ (a typical material easily microwave sinterable with NTC behavior of its electric resistivity). The Ti-6Al-4V sample sensible to the magnetic field is located at the waveguide edge (maximum magnetic field). The ceramic sample sensible to the electric field is located in the waveguide center (maximum electric field). The temperature dependences at 50% of the relative density for the main contributing dissipation terms (permeability imaginary part for the alloy and permittivity imaginary part of the ceramic) are plotted in fig.6a. The corresponding temperature evolution curves for the constant input microwave power of 2000 W are reported in fig.6b. Concerning the ceramic sample, the dissipative term is of a low magnitude at room temperature and is strongly increased at high temperature. This permittivity imaginary part generates a characteristic thermal runaway profile under microwave heating [15,16]. On the contrary, the alloy's dissipative term decreases slowly with the temperature and generates a heating profile where the temperature rapidly increases in the beginning and is stabilized at about 700 K. The difference between 3Y-ZrO$_2$ and Ti-6Al-4V heating profiles under microwave illumination is that the alloy does not render a thermal runaway, and the temperature field of the sample is more homogeneous compared to the ceramic specimen where a hot spot is generated [15]. In this simple waveguide simulation, a TE10 port condition is used without any other tooling like susceptor or thermal insulator. The heating of the sample is then not disturbed by any external factor within the heating balance or any potential resonance phenomena. In this type of simulation, the sample heating depends mainly on the sample electro-magnetic properties. Therefore, the ultrafast heating experimentally observed for Ti-6Al-4V cannot be explained by the alloy properties.

To explain this phenomenon, a fully-coupled EMTM simulation including all the tooling materials and the real cavity geometry is carried out. The results showing the electric, magnetic, thermal and sintering densification fields are reported in fig.7 (an animation is also provided in the supplementary material). The calculated evolution of the physical parameters indicates two main process stages. During the first stage, the alloy powder heating has a usual behavior with an initial fast heating followed by stabilization. During this first stage, all the other physical parameters ($T$ temperature, $E$ electric field, $H$ magnetic field, $D$ relative density) are stable. At the onset temperature of 800°C, the second stage starts where all the parameter magnitudes ($E, H, T, D$) start to increase. The sample temperature profiles show an unexpected runaway accompanied by a very fast sintering. The sintering time is about 60 s and it takes place during the main temperature runaway stages where the heating rate is about 700 K/min up to a temperature close to the melting point.

The reflective power coefficient S11 profile reported in fig.8 indicates a resonance pick during the thermal runaway. The resonance phenomena depend on many parameters such as frequency, cavity dimensions, and material electromagnetic properties inside the cavity. The electric field distribution during the runaway stage shown in fig.7 indicates a very high magnitude in the zirconia ball bed. At the beginning of the cavity resonance phenomenon, the temperature of the zirconia bed corresponds to the onset temperature of the zirconia permittivity imaginary part shown in fig.6a. The consequence of this resonance is a significant increase of the magnetic field intensity (fig.7) that drastically increases the sample temperature (sensible to the magnetic field). Since the sample is surrounded by the nano-SiC powder whose thermal conductivity is very low, the heat dissipated in the sample is confined in it, which also contributes to the acceleration of the sample's heating rate. The sample thermal runway is the result of the combination of multiple factors including the resonance phenomenon, the location of the sample in a high magnetic field area and the confinement of the heat generated in the sample by the nano-SiC powder.

The originality of this microwave flash sintering approach is in the presence of a metal thermal runaway which originates not from the metal properties but from the cavity resonance activated by the surrounding tooling material (here - zirconia). This phenomenon observed for microwave sintering is similar to the known flash sintering phenomena in terms of the sintering time and temperature profile and it is now extended to metals.

Based on the above-mentioned analysis, one can formulate the methodology of using microwave heating with favorable conditions allowing the metal flash phenomena. In particular, EMTM simulation allows a parametric study able to predict the tooling temperature-dependent properties' impact on the cavity resonance and the high magnetic fields' locations. In this connection the following principles can be formulated:

1. The microwave cavity and the temperature dependent material properties in it have to favor the appearance of the resonance phenomenon.

2. The onset time of the resonance phenomenon is a crucial point. Under constant microwave power the temperature dependence of the tooling electromagnetic properties tends to activate a resonance phenomenon like the one shown in fig.8, which is brief and quickly destroyed. If the resonance pick appears too early, the thermal runaway of the sample can be stopped before the temperature is high enough to quickly sinter the sample. In the case considered in the present study, the maximum temperature is 1595°C, which is very close to the Ti-6Al-4V sample melting point of 1604°C. To modify the resonance pick onset time, it is possible to adjust the tooling configuration inside the cavity or the cavity dimensions itself.

3. The last point is to ensure that the metallic powdered sample is located in the maximum magnetic field area and is surrounded by a highly thermo-insulating tooling. This approach allows sintering simultaneously several small samples or a medium size sample (no more than 50 mm considering the wave length at 2.45 Ghz) in the high intensity magnetic field area.

*4.5 Sample shape distortions, and gradients origin*

The consolidated sample has a truncated-cone shape with the top area more densified than the bottom area. The magnetic field, temperature, and relative density distribution at the end of the thermal runaway are reported in fig.9. The magnetic field distribution includes two main high magnitude areas. The sample is in contact from its top face with the upper cavity high magnetic field area. The top sample face is therefore a hot spot area that generates a densification gradient where the top part of the sample is more densified than the bottom part. In accordance with the experiment described in fig.5, a small truncated-cone shape distortion is related to the densification gradient.

*4.6 Microwave vs conventional sintering comparison, evidence of sintering kinetic acceleration under microwave heating*

In order to compare the microwave and conventional sintering behaviors, the dilatometer test has been carried out; it is reported in fig.10a. The conventional sintering densification behavior at 10K/min has been determined and is reported in Table A. When the determined conventional sintering kinetics is used to describe the microwave flash sintering (fig.10b), the resulting densification curve (shown by red crosses in fig.10c) exhibits the densification of only 1%. The conventional sintering kinetics is therefore significantly slower than the kinetics of the densification under microwave flash conditions. To model the microwave sintering kinetics and to explain accurately the experimentally observed sample distortions, a modified sintering model has been assumed (blue color curve in fig.10c, the respective equations are given in Table A). The sintering kinetics is then drastically accelerated under microwave flash heating conditions. Different possible reasons may explain this phenomenon: the microwave induced ponderomotive forces [38,39], the high heating rate impact on the porosity [40], the particle surface softening/melting [13]; it is also possible that the sintering kinetics has a specific behavior when the specimen's temperature is close to the melting point.

## 5. Conclusions

Ti-6Al-4V powdered specimens have been flash sintered by microwave heating. The thermal runaway accompanied by very fast sintering (60 s) is experimentally observed. An electromagnetic-thermo-mechanical simulation able to predict the microwave field distribution in the cavity, the temperature field, and the sample densification has been employed to explain the origin of the flash sintering phenomenon. The model reveals that the thermal runaway originates not from the alloy electromagnetic properties but from the resonance phenomenon activated by the surrounding tooling (zirconia) NTC behavior. The simulation successfully explains the sample shape distortions and densification gradients observed experimentally. The densification gradient is explained by the wave distribution in the cavity that generates a hot spot at the top part of the sample. Except for the fast processing time, the advantages of this microwave flash approach are:

- no need of preheating in a furnace;
- the process is contactless and does not require electric connections;
- the sample is heated in volume and is thermally confined by the nano-SiC powder;
- this microwave approach allows an extention of the flash sintering phenomenon to non-NTC materials like metals and alloys.

The last point is important as nearly no metal flash sintering with thermal runaway heating profiles is reported in the literature. Overall, the present study demonstrated the possibility of the control of the microwave flash sintering in terms of the temperature/relative density gradients which, in turn, influence the homogeneity of the sintered microstructures.

## Acknowledgements

The authors gratefully acknowledge the support from the Office of Naval Research, Contract # N00014-14-C-0233, and Dr. William Mullins, Program Officer.

**Figure captions**

Fig.1  Microwave furnace schematics a) cavity, different sample heating modes b) free surface in the center c) free surface on the edge d) nano-SiC powder surrounded sample.

Fig.2  Ti-6Al-4V powder, a) SEM image, b) EDS analysis.

Fig.3  Ti-6Al-4V powder complex permittivity and permeability extraction.

Fig.4  a) free surface sample heating mode b) nano-SiC powder-surrounded sample's heating mode.

Fig.5  Etched microstructure of the flash sintered sample.

Fig.6  Ceramic/alloy microwave heating comparison: a) temperature dependence of the most relevant dissipative terms, b) heating curves under 2000 W microwave power and samples' location in the waveguide.

Fig.7 Electromagnetic thermo-mechanical simulation of the flash sintering process, visualization of the electric, magnetic, thermal fields and the sample relative density evolution.

Fig.8  Cavity resonance profile: a) reflective coefficient b) sample temperature.

Fig.9  Sample gradients and distortions investigation at the end of the sintering, a) simulated magnetic field, b) simulated temperature field, c) experimental microstructures and sample shape, d) simulated relative density field.

Fig.10 a) experimental/modeled dilatometer relative density curve, b) microwave flash sintering sample temperature curve, c) experimental/ modeled microwave flash sintering relative density curves (with the identified dilatometer behavior and microwave modified behavior).

**Table captions (Appendix)**

Table A: Electromagnetic-thermal-mechanical materials properties; with T the absolute temperature, D the relative density, $Cp$ (J .kg$^{-1}$.K$^{-1}$) the specific heat, $\kappa$ (W.m$^{-1}$.K$^{-1}$) the thermal conductivity, $\rho$ (kg .m$^{-3}$) the density, $\epsilon$ the emissivity, $\varepsilon'_r$, $\varepsilon''_r$, $\mu'_r$, $\mu''_r$ the permittivity and permeability real and imaginary parts respectively and R the gas constant, A the sintering viscosity (s.Pa).

# Figures:

Fig.1 Microwave furnace schematics a) cavity, different sample heating modes b) free surface in the center c) free surface on the edge d) nano-SiC powder surrounded sample.

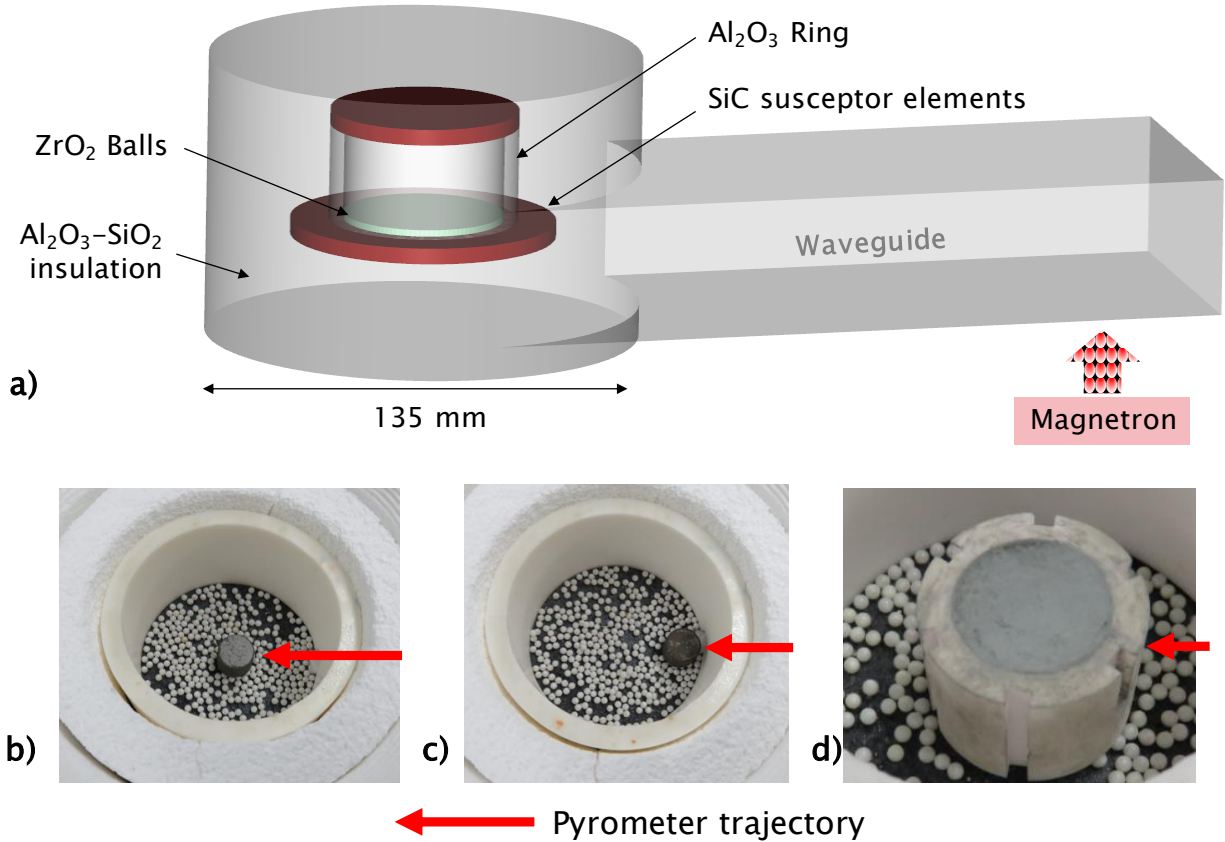

Fig.2 Ti-6Al-4V powder, a) SEM image, b) EDS analysis.

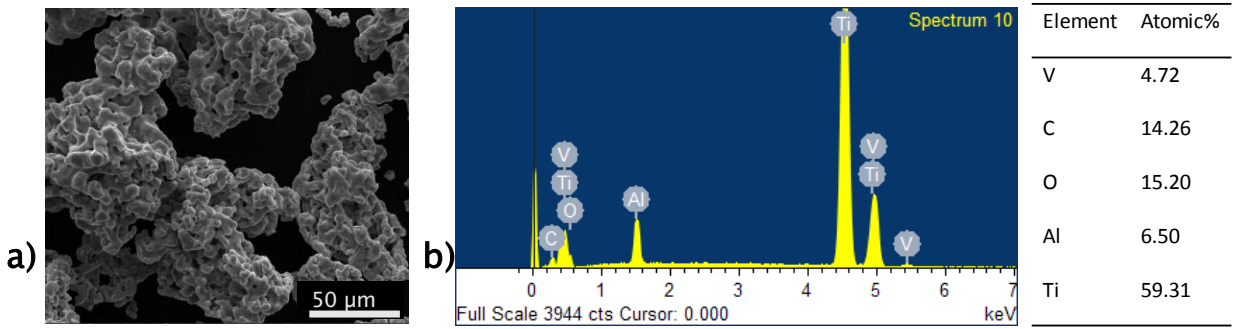

Fig.3 Ti-6Al-4V powder complex permittivity and permeability extraction.

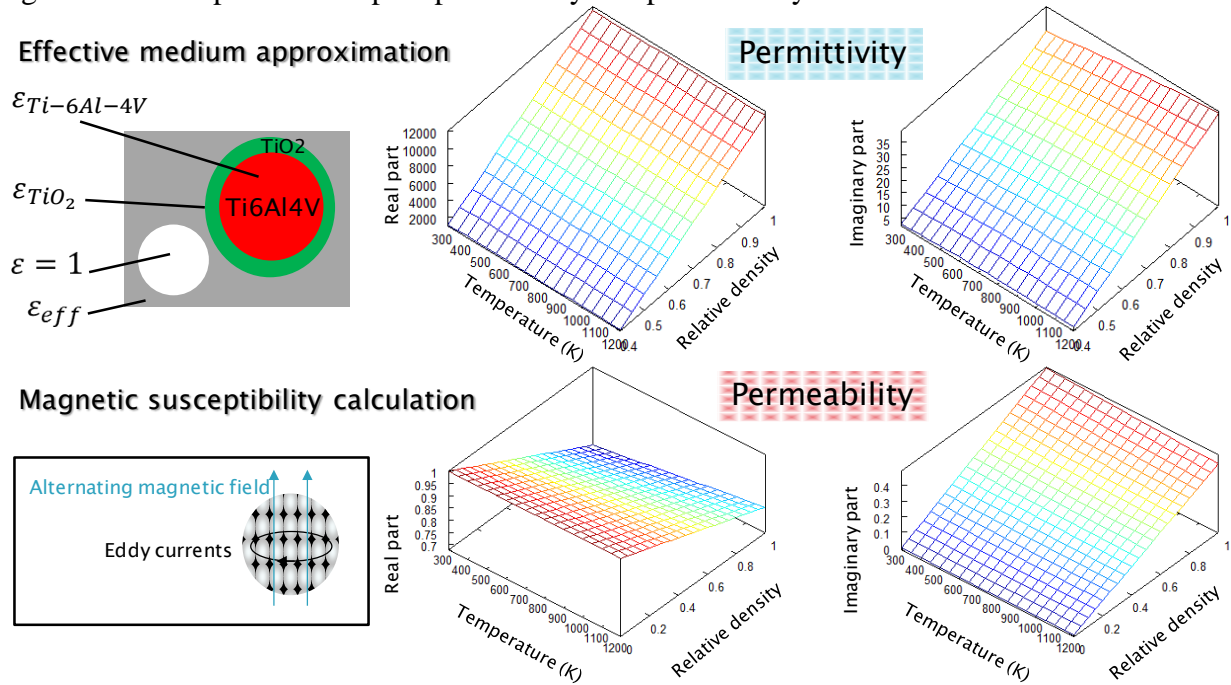

Fig.4 a) free surface sample heating mode b) nano-SiC powder surrounded sample's heating mode.

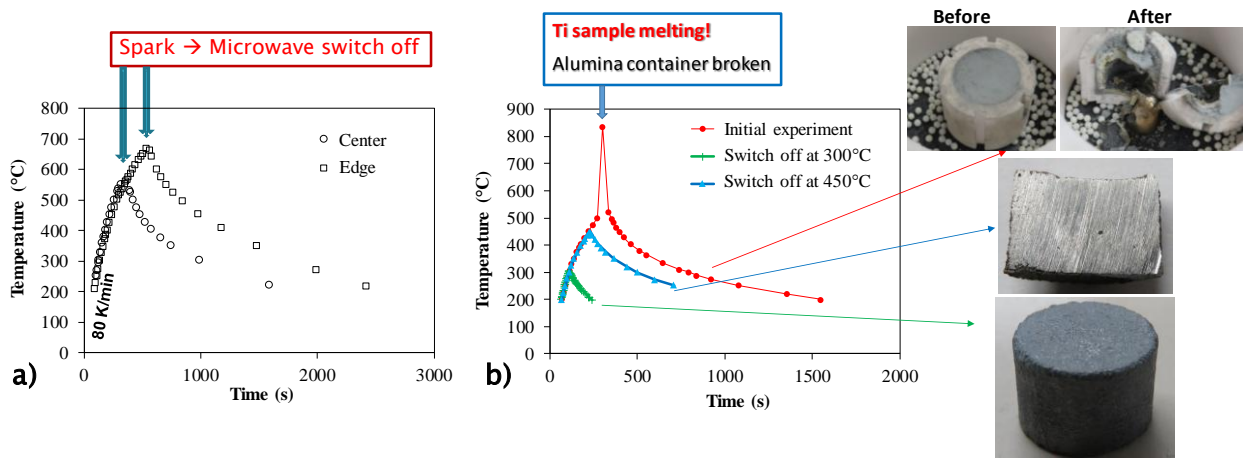

Fig.5 Etched microstructure of the flash sintered sample.

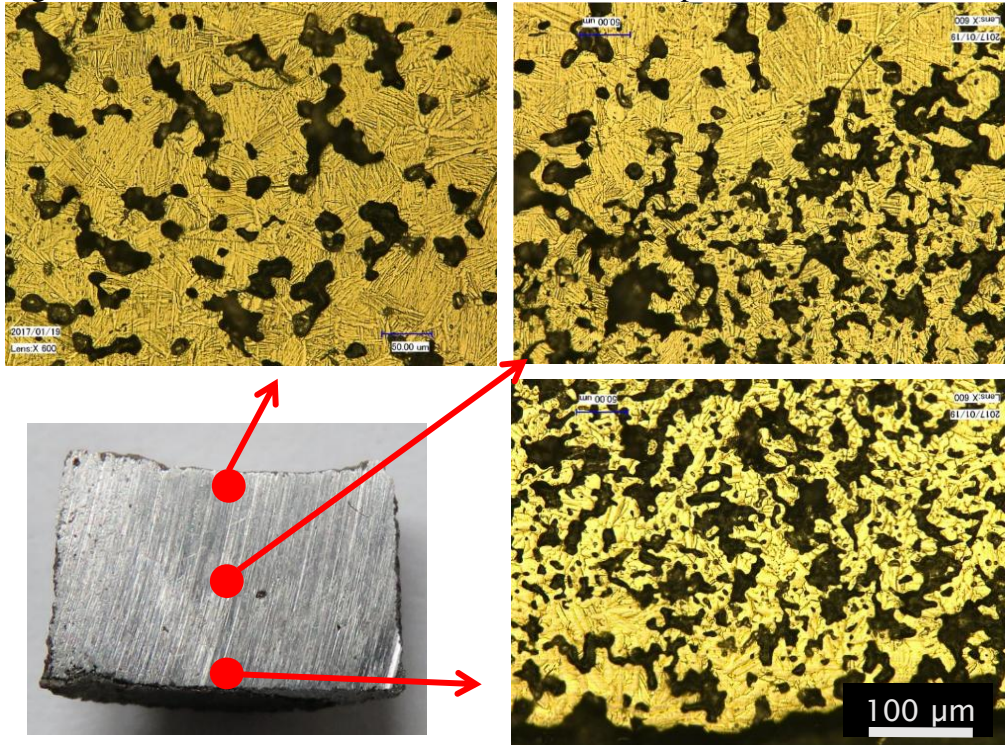

Fig.6 Ceramic/alloy microwave heating comparison: a) temperature dependence of the most relevant dissipative terms, b) heating curves under 2000 W microwave power and samples' location in the waveguide.

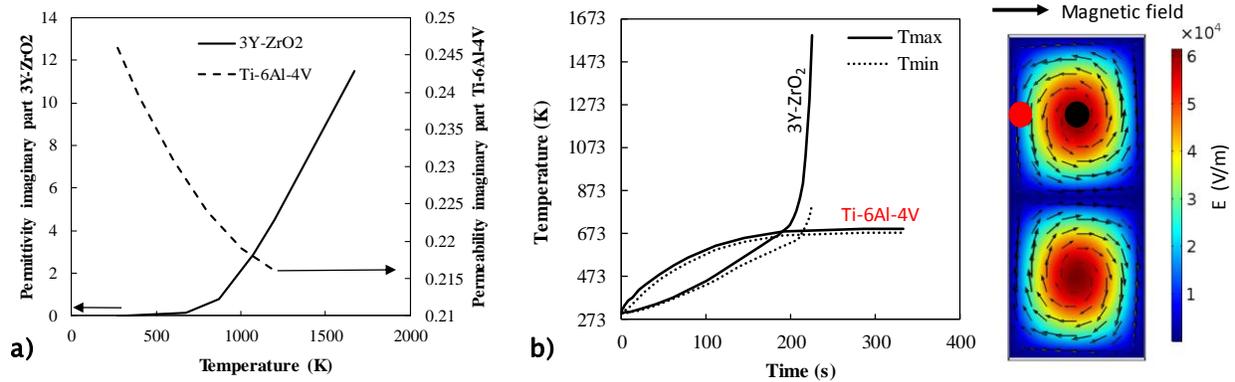

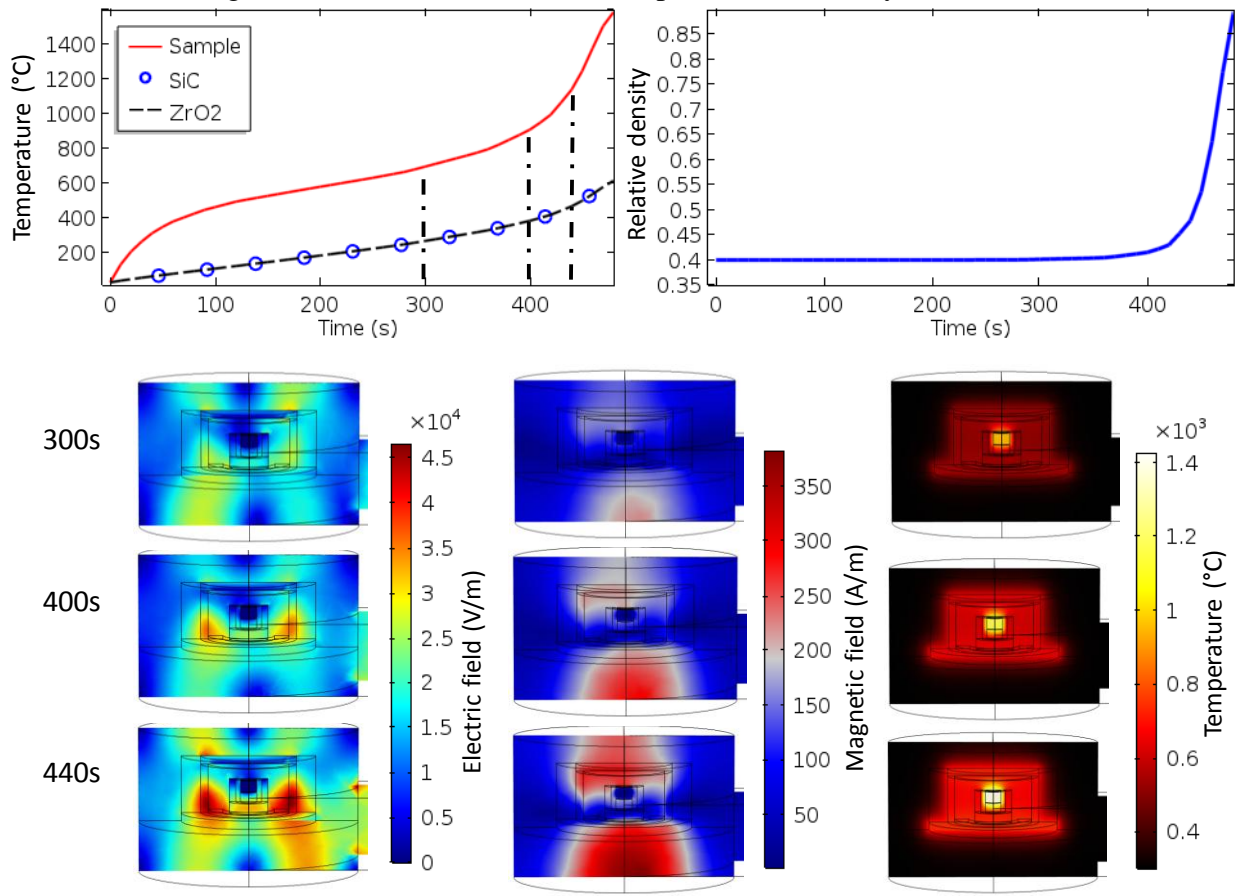

Fig.7 Electromagnetic thermo-mechanical simulation of the flash sintering process, visualization of the electric, magnetic, thermal fields and the sample relative density evolution.

Fig.8 Cavity resonance profile: a) reflective coefficient b) sample temperature.

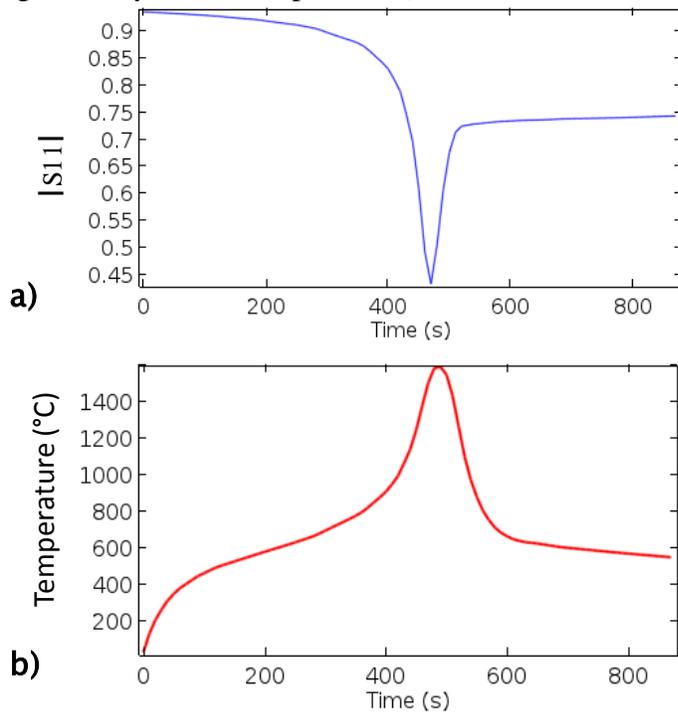

Fig.9 Sample gradients and distortions investigation at the end of the sintering, a) simulated magnetic field, b) simulated temperature field, c) experimental microstructures and sample shape, d) simulated relative density field.

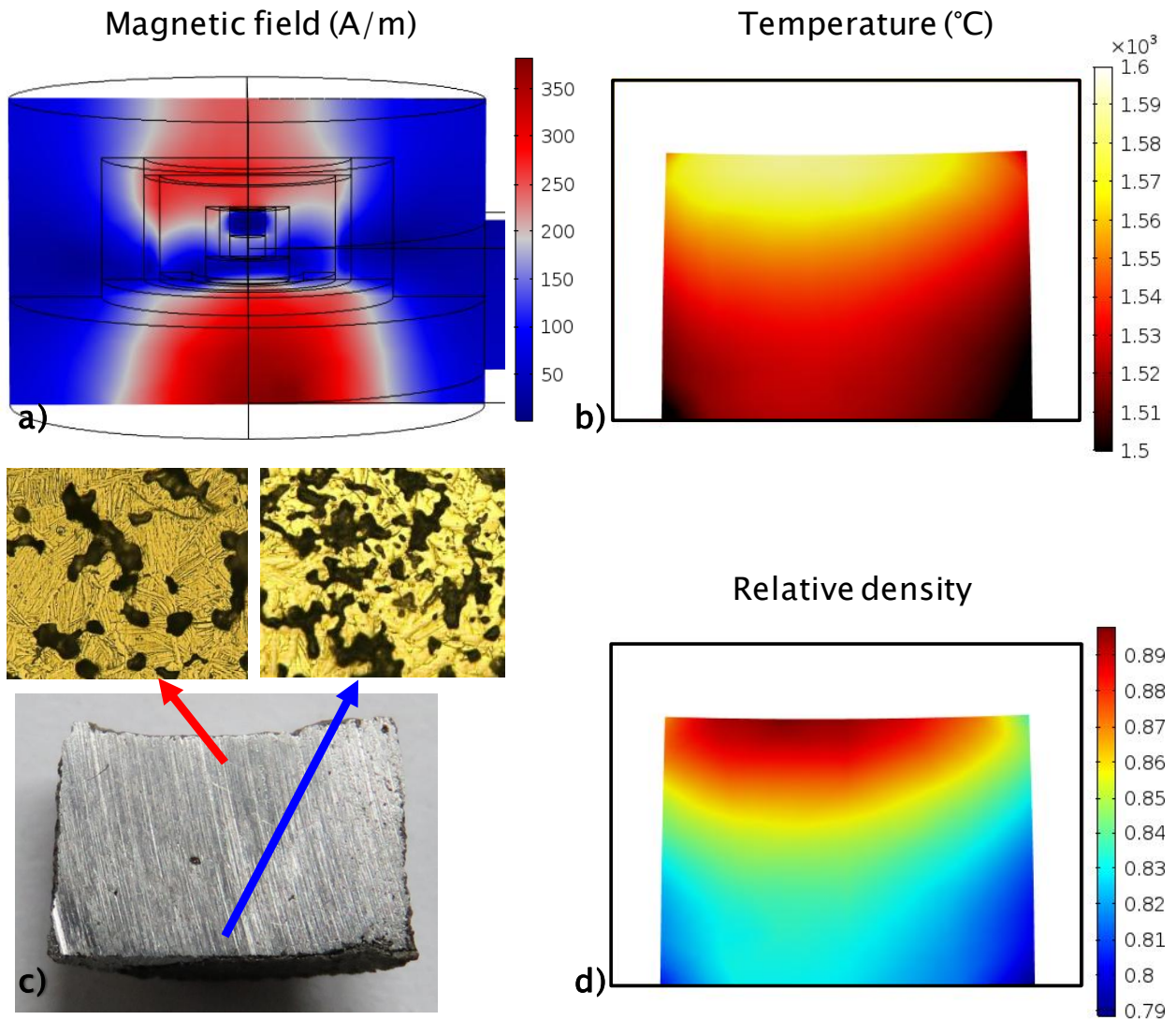

Fig.10 a) experimental/modeled dilatometer relative density curve, b) microwave flash sintering sample temperature curve, c) experimental/ modeled microwave flash sintering relative density curves (with the identified dilatometer behavior and microwave modified behavior).

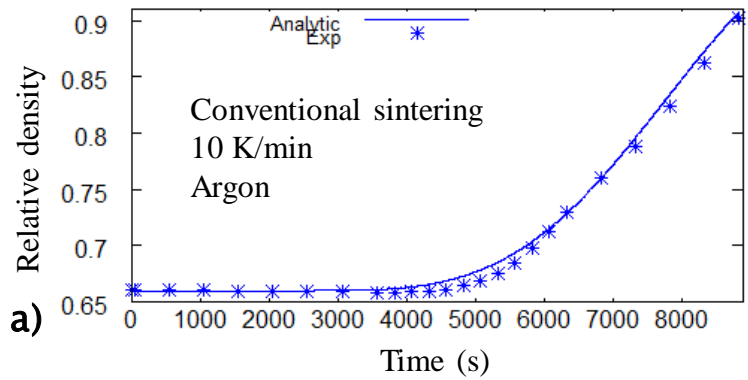

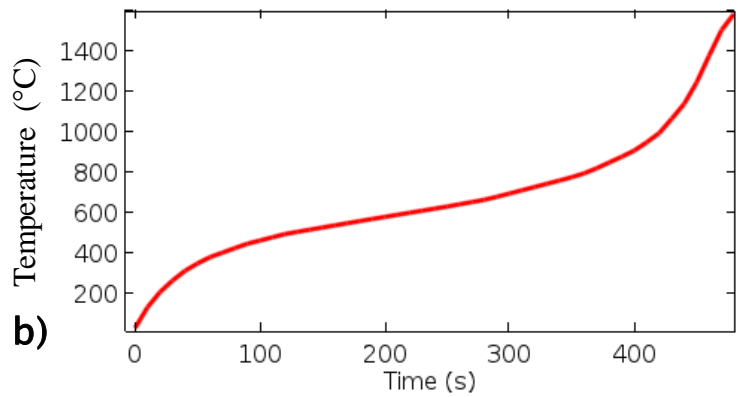

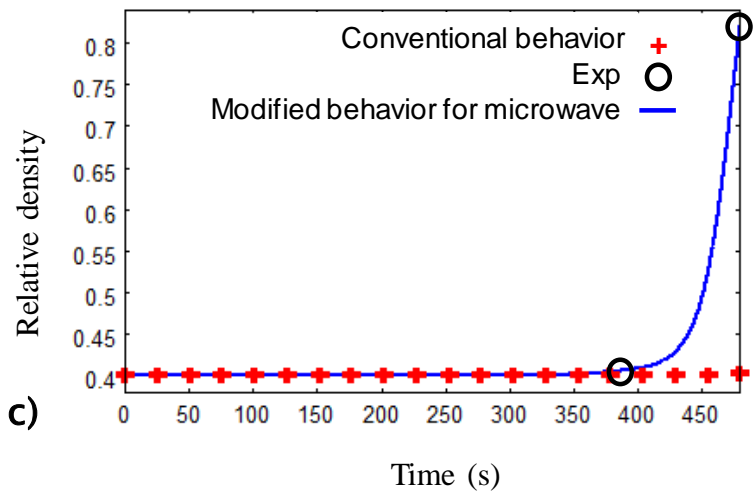

Table A: Electromagnetic-thermal-mechanical materials properties; with T the absolute temperature, D the relative density, $Cp$ (J .kg$^{-1}$.K$^{-1}$) the specific heat, $\kappa$ (W.m$^{-1}$.K$^{-1}$) the thermal conductivity, $\rho$ (kg .m$^{-3}$) the density, $\epsilon$ the emissivity, $\varepsilon'_r$, $\varepsilon''_r$, $\mu'_r$, $\mu''_r$ the permittivity and permeability real and imaginary part respectively and R the gas constant, A the sintering viscosity (s.Pa) **(Appendix A).**

| | Temperature range (K) | Zirconia bed [15,24] | Temperature range (K) | Silicon carbide susceptor [24,31] | nanoSiC powder [24,31] | Temperature range (K) | 80% Al2O3-20% SiO2 [16] | Al$_2$O$_3$ [28–30] | Temperature range (K) | Ti-6Al-4V powder [32] |
|---|---|---|---|---|---|---|---|---|---|---|
| $Cp$ | 273-1473 | (43+2.35 T-0.34E-3 T$^2$+4.25E-6 T$^3$-2.09E-9 T$^4$+4.06E-13 T$^5$) × D | 273-673 | -8.35+3.08T-0.00293 T$^2$+1.0268E-6 T$^3$ | (-8.35+3.08T-0.00293 T$^2$+1.0268E-6 T$^3$) ×0.048 | 273-2200 | -5.31E-4.T$^2$+1.25.T+5.18E2 | 850 | 273-1873 | (383+0.671.T-5.35E-4.T$^2$+1.64E-7.T$^3$)×D |
| | | | 673-1573 | 772+0.431 T-2.10E-5 T$^2$ | (772+0.431 T-2.10E-5 T$^2$) ×0.048 | | | | | |
| | 1473-2200 | 638 × D | 1573-2200 | 1400 | 1400×0.048 | | | | | |
| $\kappa$ | 273-2200 | (1.96-2.32E-4 T+6.33E-7 T$^2$-1.91E-10 T$^3$) × (1-1.5 × (1-D)) | 273-2200 | 192-0.326 T+2.74E-4 T$^2$-7.71E-8 T$^3$ | 6.61E-5.T+1.31E-2 | 273-2200 | 1.40E-4.T+1.70E-2 | 39500.T$^{-1.26}$ | 273-1873 | (8.11-0.0149.T+4.47E-5.T$^2$-2.27E-8.T$^3$) × (1-1.5 × (1-D)) |
| $\rho$ | 273-2200 | (6132 -9.23E-2 T-7.26E-5 T$^2$+4.58E-8 T$^3$-1.31E-11 T$^4$) × D | 273-2200 | 2977+0.0510 T-2.29E-4 T$^2$+2.98E-7 T$^3$-1.92E-10 T$^4$+4.77E-14 T$^5$ | (2977+0.0510 T-2.29E-4 T$^2$+2.98E-7 T$^3$-1.92E-10 T$^4$+4.77E-14 T$^5$) ×0.048 | 273-2200 | -1.04E-02.T+4.43E2 | 3899 | 273-1873 | (4467-0.119.T-1.28E-5.T$^2$) ×D |
| $\epsilon$ [27] | 273-2200 | 0.7 | 273-2200 | 0.9 | 0.9 | 273-2200 | 0.83 | 0.8 | 273-1873 | - nano-SiC powder surrounded |
| $\varepsilon'_r$ | 273-2200 | -5.38-4.34E-3 T+2.22E1 D+1.37E-2 T D | 273-2200 | 1.88E-06 T$^2$-1.67E-03 T+6.4 | 1.26E-8.T$^2$ - 1.03E-5.T+ 1.11 | 273-2200 | 5.03E-8.T$^2$+1.37E-5.T+1.5 | 1.34E-3.T+8.3 | 273-1873 | 5744+0.0000298.T+17200.D-0.0000893.T.D |
| $\varepsilon''_r$ | 273-673 | 1.48E-1-5.76E-4 T-4.55E-01 D+1.77E-03 T D | 273-2200 | 2.36E-12 T$^4$-7.15E-09 T$^3$+7.72E-06 T$^2$-3.43E-03 T+9.92E-01 | 4.47E-15.T$^4$ - 9.08E-12.T$^3$ + 4.37E-9.T$^2$ + 6.23E-7.T + 3.25E-3 | 273-2200 | 2.50E-09.T$^2$-1.64E-6.T+3.96E-4 | 4.62E-05.T-8.87E-3 | 273-1873 | -15.0+0.0009908.T+39.6.D+0.0153.T.D-0.00000406.T$^2$+0.000323.D$^2$ |
| | 673-873 | 3.82-6.03E-3 T-1.172E1 D+1.85E-2 T D | | | | | | | | |
| | 873-1073 | 1.56E1-1.95E-2 T-4.74E1 D+5.94E-2 T D | | | | | | | | |
| | 1073-2200 | 3.25E1-3.86E-2 T-7.64E1 D+8.46E-2 T D+3.82E-6 T$^2$+1.07 D$^2$ | | | | | | | | |
| $\mu'_r$ | 273-2200 | 1 | 273-2200 | 1 | 1 | 273-2200 | 1 | 1 | 273-1873 | 1+7.43E-18.T-3.52E-1.D+1.14E-4.T.D |
| $\mu''_r$ | 273-2200 | 0 | 273-2200 | 0 | 0 | 273-2200 | 0 | 0 | 273-1873 | 1.03E-2-3.42E-5.T+5.04E-1.D-6.34E-5.T.D+2.28E-8.T$^2$+1.40E-16.D$^2$ |
| A | 273-2200 | - | 273-2200 | - | - | 273-2200 | - | - | 273-1873 | 6802.T exp(70000/(RT)) Dilatometer behavior 154.T exp(70000/(RT)) Microwave behavior |